\documentclass[twocolumn,prb,showpacs]{revtex4}

\usepackage{bbold,epsfig,amsmath,amssymb,color,graphicx}
\begin{document}

\title{Unveiling the internal entanglement structure of the Kondo singlet}
\author{Chun Yang}
\affiliation{Department of Physics, Northeastern University, Boston, Massachusetts 02115, USA}
\author{Adrian E. Feiguin}
\affiliation{Department of Physics, Northeastern University, Boston, Massachusetts 02115, USA}

\date{\today}
\begin{abstract}
We disentangle all the individual degrees of freedom in the quantum impurity problem to deconstruct the Kondo singlet, both in real and energy space, by studying the contribution of each individual free electron eigenstate. This is a problem of two spins coupled to a bath, where the bath is formed by the remaining conduction electrons. Being a mixed state, we resort to the ``concurrence'' to quantify entanglement. We identify ``projected natural orbitals'' that allow us to individualize a single-particle electronic wave function that is responsible of more than $90\%$ of the impurity screening. 
In the weak coupling regime, the impurity is entangled to an electron at the Fermi level, while in the strong coupling regime, the impurity counterintuitively entangles mostly with the high energy electrons and disentangles completely from the low-energy states carving a ``hole'' around the Fermi level. This enables one to use concurrence as a pseudo order parameter to compute the characteristic ``size'' of the Kondo cloud, beyond which electrons are are weakly correlated to the impurity and are dominated by the physics of the boundary. 
\end{abstract}
\maketitle

\section{Introduction}
The Kondo problem describes a magnetic impurity screened by the spin of the electrons in the Fermi sea,
forming a collective singlet state\cite{HewsonBook,cox1998exotic}.
Its simplest formulation is through the so-called Kondo impurity model:
\begin{equation}
H =  \sum_{k\sigma} \epsilon_k c_{k\sigma}^\dagger c_{k\sigma} + J_K \vec{S}_{\rm imp } \cdot \vec{S}_{\rm r_0},
\label{hamiltonian}
\end{equation}
where the $\vec{S}_{\rm imp}$ operator represents an $S=1/2$ impurity embedded in a Fermi sea of non-interacting fermions. The Kondo interaction with a fermion at position $r_0$ is parametrized by the coupling $J_K$.
Irrespective of the dispersion $\epsilon_k$, the problem is intrinsically one-dimensional and several approaches, such as the Numerical Renormalization Group (NRG)\cite{nrg_wilson,nrg_bulla}, the Bethe Ansatz \cite{andrei1983solution,tsvelik1983I}, and generalizations to lattice problems \cite{busser2013lanczos}, take advantage of this low dimensionality. 

Most of our understanding of the Kondo problem stems from renormalization group (RG) formalisms that yield a complete physical picture of the distance independent physics in the strong coupling limit and at low energies, as zooming out from the impurity and looking at it from afar \cite{Anderson1970}. 
This fixed point is characterized by a single energy scale -- the Kondo temperature $T_K$ -- and is described by a bound state formed by the impurity and the conduction electrons, the ``Kondo singlet''.
 This wave function is typically characterized as a screening cloud (``Kondo cloud'') centered at the impurity and decaying in distance with a characteristic range $\xi_K$ \cite{Sorensen1996,affleck2001detecting,Sorensen2005a,affleck2009kondo,bergmann08,FHM09,Busser10} 
that depends on $J_K$ (or $T_K$). At distances of the order of $\xi_K$, or in finite systems, where the conduction electrons are confined to a small spatial region in a ``Kondo box''\cite{schlottmann2001kondo,simon2002finite,simon2003kondo,kaul2006spectroscopy,hand2006spin} {\it all} the electrons may be inside the Kondo cloud, without an {\it outside}. This regime would correspond to the ``crossover'' between weak coupling and strong coupling in the RG flow.

In the strong coupling limit for $J_K$ much larger than the bandwidth $W$, it is easy to visualize a tightly bound singlet formed by the the impurity and a localized electron at $r_0$. Nozi\`eres elegantly demonstrated\cite{Nozieres1976,Nozieres1980} that this fixed point can be described within Fermi liquid theory: the bound state becomes just a scattering center and the remaining conduction electrons that are not coupled to the impurity will simply behave as free fermions with their wave functions modified by a phase shift $\delta=\pi/2$.
At intermediate couplings, as we reduce $J_K$, the impurity will become correlated with electrons farther and farther from it. In a small system, at some point the Kondo cloud will not fit into the ``box'' and it cannot form: the singlet will extend to the entire volume and the impurity will couple mostly to one electron at the Fermi level.

Even though this is one of the most studied and best understood problems in condensed matter physics, deeply conflicting pictures coexist when it comes to interpreting the internal structure of this state. 
For instance, how is it possible that in a dilute system with numerous impurities with overlapping Kondo clouds, Kondo physics dominates and a single impurity model can explain all experimental observations?\cite{Coleman2002,affleck2009kondo} In the case of a large number of impurities, one would expect that at some point there would not be enough electrons near the Fermi surface to screen all the impurities, as postulated by Nozi\'eres in his ``exhaustion'' paradox\cite{Nozieres1985,Nozieres1998}. 
Understanding the screening process and the internal entanglement structure of the Kondo singlet is paramount to understanding more complex problems, such as ``exhaustion'' in heavy-fermion systems\cite{Nozieres1985,Nozieres1998} and the so-called ``Kondo breakdown'' \cite{Coleman2001,Hackl2008}. In this work we make significant progress in this direction by using quantum information ideas to quantify the two-particle entanglement between the impurity and each conduction electron individually.



The paper is organized as follows: in Section II we describe the model and methods utilized in the calculations, in Section III we introduce the definition of entanglement in a mixed state of two spins and the idea of ``concurrence'' and in Section IV we show results and the analysis of the pairwise entanglement. In Section V we introduce the concept of projected natural orbitals and we discuss their relevance in terms of representing the internal entanglement structure of the wave-function. Based on this, in Section VI we show how entanglement can be used to estimate the Kondo screening length. We close with a summary and conclusions. 

\begin{figure}
 \includegraphics[height=0.48\textwidth, angle=-90]{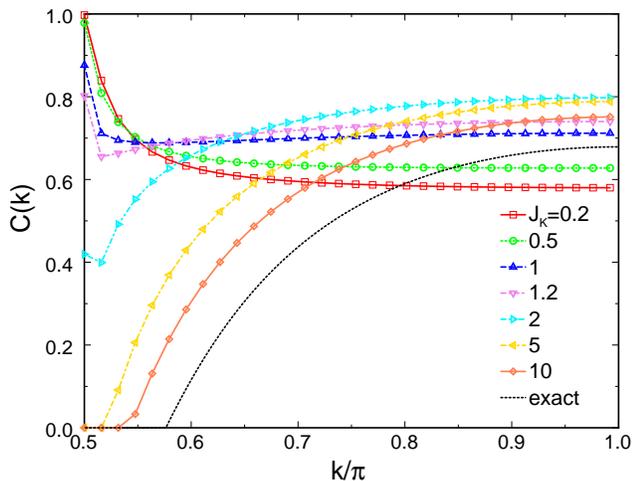}
\caption{Concurrence momentum distribution $C(k)$ between a Kondo impurity and the electrons in the Fermi sea for $N=L+1=63$. The exact results correspond to the limit $J_K \rightarrow \infty$, $L \rightarrow \infty$. $C(k)$ is symmetric about $k_F=\pi/2$.
} 
\label{fig:concurrence}
\end{figure}


\section{Model and methods}
Without loss of generality, we model the conduction electrons by a one dimensional tight-binding chain with open boundary conditions of size $2L+1$ ($L$ even) and the impurity connected to the site in the middle\cite{Sorensen1996}. A simple folding transformation\cite{feiguin2011reducing,busser2012designing} allows us to map it onto an equivalent chain of length $L+1$ and an impurity coupled to the first site ($r_0=0$):
\begin{equation}
H_{\rm el} = -\sqrt{2}t \sum_{\sigma} (c_{0\sigma}^\dagger c_{1\sigma}+\mathrm{h.c.}) - t\sum_{\sigma,i=1}^{L-1} (c_{i\sigma}^\dagger c_{i+1\sigma}+\mathrm{h.c.}),
\end{equation}
where $t$ is the hopping matrix element and our unit of energy.
The electrons in this Hamiltonian correspond in reality to the symmetric (bonding) electrons of the original problem. 
In the following we refer to its eigenstates as $|k\rangle=c^\dagger_k|{\rm vac}\rangle$, with energies $\epsilon_k = -2t\cos{k}$ and momenta $k=\pi/(2L+2)j$ ($j=1,3,5,\cdots,2L+1$).

The problem can readily be solved with the density matrix renormalization group (DMRG)\cite{White1992,White1993,Schollwock2005density} and indeed, it has been studied in the literature, particularly focusing on the spatial correlations \cite{Sorensen2005a,busser2013lanczos} and the bi-partite entanglement entropy \cite{sorensen2007,affleck2009}. 
We conducted DMRG simulations in systems of up to $N=L+1=63$ orbitals at half-filling, keeping the truncation error below $10^{-8}$ which translates into up to 3000 DMRG basis states for real space simulations. 

\section{Entanglement in a mixed state}
In this work we turn our attention to the entanglement and correlations between the impurity spin and the {\it individual} electronic wave functions. This is a problem of two spins embedded in a bath formed by the rest of the conduction electrons, effectively in a mixed state. This problem has attracted a great deal of interest in the quantum information community and very seldom looked at in this condensed matter context \cite{Cho2006,Oh2006} (See Ref.\onlinecite{Amico2008} for a general review). 

 Entanglement is related to the non-locality of correlations in quantum mechanics and in pure states it can be measured through the entanglement entropy by means of the Schmidt decomposition. 
In the case of mixed states, these ideas do not apply\cite{horodecki2009}.

In order to determine whether the spins are entangled in a mixed state one needs to work with the two-particle density matrix.
In general, a density matrix is said to be separable if it can be written as
\[
\rho = \sum_i p_i \rho_{1i} \otimes \rho_{2i}.
\]
The term ``entangled'' refers to non-separable states. There are many alternative ways to define the entanglement of formation\cite{Vedral1997,Hill1997,Wootters1998,Vedral1998,Verstraete2001,Verstraete2003,Modi2010}. 
A conventional way to introduce it is as 
\cite{Bennett1996, Hill1997,Wootters1998}:
\begin{equation}
E=\min_{\{p_j,|\psi_j\rangle\}}\sum_j p_j S(|\psi_j\rangle),
\label{entropy}
\end{equation}
where the minimization is over all pure state decompositions $\rho = \sum_j p_j|\psi_j\rangle \langle \psi_j|$, and $S=-\rho_{1j}\log_2(\rho_{1j})$ is the von Neumann entropy obtained by tracing over one of the subsystems, $\rho_{1j}=\mathrm{tr}_2(|\psi_j\rangle \langle \psi_j|)$.

There is no simple solution for the separability problem, which in general is NP-hard \cite{Gharibian2010}. However, for the particular case that we study here, this is actually much simpler.
The separability of a two-spin system such as ours can be determined by means of the Peres Horodecki (PPT) criterion \cite{Peres1996}, which tells us that the necessary and sufficient condition is that the partial transpose of $\rho$ with respect to spin ``$2$'' has non-negative eigenvalues. 

In a seminal paper, Hill and Wooters found a closed formula for the entanglement of formation for two spins\cite{Hill1997,Wootters1998,Wootters2001,Amico2008}:
\[
E(\rho) = h\left( \frac{1+\sqrt{1-C(\rho)^2}}{2}\right),
\]
with
\[
h(x)=-x \log_2x-(1-x)\log_2(1-x).
\]
The quantity $C(\rho)$ is called the ``concurrence'' and has the property that it is an entanglement monotone and is zero for a separable state. For a mixed state of two spins/qubits it takes the form:
\[
C(\rho) = \max{(0,\lambda_1-\lambda_2-\lambda_3-\lambda_4)},
\]
where the $\lambda_i$'s are the square roots of the eigenvalues of the matrix $\rho \tilde \rho$ in decreasing order, with
\[
\tilde\rho = (\sigma_y \otimes \sigma_y)\rho^*(\sigma_y \otimes \sigma_y).
\]
Here, $\rho^*$ is the complex conjugate of $\rho$ in the standard basis ${|\sigma \sigma'\rangle}$.

\section{Concurrence in the Kondo problem}

A good point to start is by writing the (normalized) ground state wave function in a form that takes into account the symmetries of the problem, isolating the contributions of the impurity spin and the orbital of interest. In the absence of magnetic fields, this acquires the form:
\begin{eqnarray}
|{\rm g.s.}\rangle &=& a_l\left(|\Uparrow\rangle |2\rangle_l |\alpha_{l,\downarrow} \rangle + |\Downarrow\rangle |2\rangle_l |\alpha_{l,\uparrow}\rangle \right) \nonumber \\
& + & b_l\left(|\Uparrow\rangle |\downarrow \rangle_l |\beta_{l,\uparrow\downarrow} \rangle + |\Downarrow\rangle |\uparrow\rangle_l |\beta_{l,\downarrow\uparrow}\rangle \right) \nonumber \\
& + & c_l\left(|\Uparrow\rangle |\uparrow \rangle_l |\delta_{l,\downarrow\downarrow} \rangle + |\Downarrow\rangle |\downarrow\rangle_l |\delta_{l,\uparrow\uparrow}\rangle \right) \nonumber \\
& + & d_l\left(|\Uparrow\rangle |0 \rangle_l |\gamma_{l,\downarrow} \rangle + |\Downarrow\rangle |0 \rangle_l |\gamma_{l,\uparrow}\rangle \right),
\label{gs2}
\end{eqnarray}
where the states $|\alpha_{l,\sigma}\rangle$,$|\beta_{l,\sigma,-\sigma}\rangle$,$|\delta_{l,\sigma\sigma}\rangle$,$|\gamma_{l,\sigma}\rangle$ do not include the single particle orbital $c^\dagger_{l\sigma}$ and contain phases (signs) that are unimportant in the following discussion. The states $|\sigma\rangle_l$, $|2\rangle_l$ and $|0\rangle_l$ indicate the occupation of the single orbital, which could be in momentum space, real space, or some other representation. The coefficients $a_l$, $b_l$, $c_l$ and $d_l$ depend on the orbital $l$ and the single particle basis, and $\langle \beta_{l,\downarrow\uparrow}|\beta_{l,\uparrow\downarrow}\rangle = (c_l^2-b_l^2)/b_l^2$.

\begin{figure}
 \includegraphics[width=0.48\textwidth, angle=0]{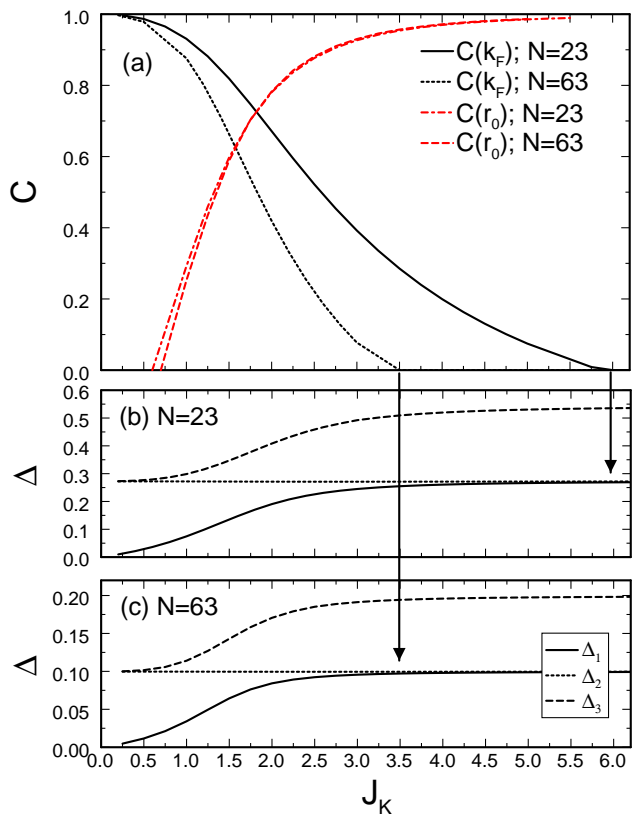}
\caption{(a) Concurrence between a Kondo impurity and the electron at momentum $k_F$ as a function of $J_K$ for system sizes $N=23$ and $N=63$. Results for the concurrence with the orbital at position $r_0$ are also shown. (b),(c) Excitation gaps above the ground state for $N=23, 63$, respectively. The arrows indicate the points at which $C(k_F)$ vanishes.
} 
\label{fig:scaling}
\end{figure}

In order to determine the entanglement between the impurity and orbital $l$, we consider that it primarily originates from the spin degree of freedom and we trace over configurations that are empty or double occupied, yielding a projected wave-function $|\phi\rangle$ not containing the terms proportional to $a_l$ and $d_l$.
This is a legitimate assumption, since the impurity only contains spin and no charge and the Kondo problem is a paradigm of spin-charge separation \cite{HewsonBook}.
In the basis ${|\Uparrow \uparrow\rangle,|\Uparrow\downarrow \rangle,|\Downarrow\uparrow\rangle,|\Downarrow\downarrow\rangle}$, both the reduced density matrix and the two-particle density matrix are identical:
\[
\rho_{\alpha \beta, \alpha' \beta'} = \frac{1}{2}\langle \phi| c^\dagger_{l\beta} c^\dagger_{\rm{imp},\alpha} c_{\rm{imp},\alpha'} c_{l\beta'}|\phi\rangle.
\]
It is a simple exercise to show that it has the form (see Appendix A):
\begin{equation}
\rho = \frac{1}{4}\left(
\begin{array}{cccc}
n_{l\uparrow}-\delta_l & 0 & 0 & 0 \\
0 & n_{l\uparrow}+\delta_l & -2\delta_l & 0 \\
0 & -2\delta_l & n_{l\uparrow}+\delta_l & 0 \\
0 & 0 & 0 & n_{l\uparrow}-\delta_l 
\end{array}
\right),
\label{rho2}
\end{equation}
with $\delta_l = -2\langle S^z_{\rm imp} n_{l\uparrow}\rangle=-2\langle S^z_{\rm imp} S^z_{l}\rangle$ (for a singlet), and all averages are with respect to the (unnormalized) state $|\phi\rangle$. Using the form of $|\phi\rangle$, these quantities can be expressed as $n_{l\uparrow}=c_l^2+b_l^2$ and $\delta_l=(b_l^2-c_l^2)$.
After dividing by $n_{l\uparrow}$ to normalize the trace, this density matrix acquires the peculiar form of a so-called ``Werner state''\cite{Werner1989,Amico2008}:
\[
\rho' = \frac{1}{4}(1-\Delta )\mathbb{1} + \Delta|s\rangle\langle s|,
\]
where $\rho'=\rho / n_{l\uparrow}$, $\Delta =\delta/ n_{l\uparrow}$ and $|s \rangle=1/\sqrt{2}(|\uparrow \downarrow \rangle- |\downarrow \uparrow \rangle)$ represents a singlet.

The PPT criterion implies that Werner states are separable for $\Delta < 1/3$. 
In addition, the concurrence is given as:
\begin{equation}
C=\max{\left(0,\frac{3\Delta-1}{2}\right)}=\max{\left(0,-\frac{\langle \vec{S}_{\rm imp}\cdot \vec{S}_{2}\rangle}{n_{2\uparrow}}-\frac{1}{2}\right)}.
\label{concurrence}
\end{equation}

\begin{figure}
 \includegraphics[height=0.48\textwidth, angle=-90]{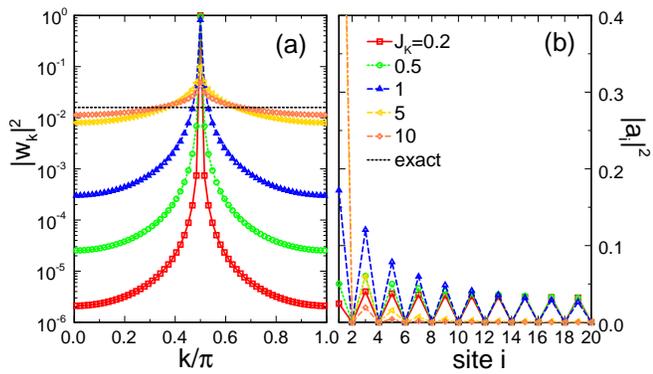}
\caption{
Wave function amplitudes of the dominant natural orbital decomposed in both momentum (a) and real space (b) components, for different values of $J_K$. For small $J_K$ the natural orbitals corresponds to the electronic wave function at $k_F$, while in the strong coupling limit, it is equal to an electron localized on the first site of the chain, or an equal superposition of all momenta.  
}
\label{fig:no}
\end{figure}

We want to evaluate the concurrence between the quantum impurity and conduction electrons $c_k$. For this purpose, we use the non-interacting form of the wave functions (without the impurity) $c_k = \sum_i U_{k i} c_i$ and calculate the correlations between the localized spin and these orbitals that yield the coefficients of the wave function (\ref{gs2}) in momentum space: $a_k=\langle n_{\rm{imp},\uparrow} n_{k\uparrow} n_{k\downarrow}\rangle$, $b_k=\langle n_{\rm{imp},\uparrow} n_{k\downarrow} (1-n_{k\uparrow})\rangle$, $c_k=\langle n_{\rm{imp},\uparrow} n_{k\uparrow} (1-n_{k\downarrow})\rangle$. 
As we discuss below, the simulation can be more efficient if carried out directly in momentum space.

We shall call the results the ``concurrence distribution'' to refer to the dispersion in momentum or energy of this quantity. Results for $C$ (for $N=L+1=63$ electrons) are plotted in Fig.\ref{fig:concurrence} for different values of $J_K$ and show a clear and dramatic change of behavior for weak and strong coupling: For small $J_K$ the impurity is entangled mostly to a single electron at the Fermi energy. As $J_K$ increases, it becomes entangled to higher energy electrons and the entanglement with the electron at $\epsilon_F$ is continuously suppressed. Eventually, in the strong coupling regime, the impurity couples mostly to high energy electrons and decouples {\it completely} from the electron at the Fermi level. 
 There is a broad range of momenta around the Fermi level where $C$ is identically zero and in the figure, it looks as though the impurity has carved a hole in the concurrence distribution.
 Using the exact solution for $J_K \rightarrow \infty$ and $L \rightarrow \infty$ one can readily verify that in this limit $a_{k_F}^2=b_{k_F}^2=c_{k_F}^2=d_{k_F}^2=1/8$, which yields $C=0$.
The behavior of $C(k_F)$ is also shown in Fig.\ref{fig:scaling}(a) for two system sizes.

Remarkably, the concurrence in real-space is always zero except for the first site of the chain $r_0$ (also shown in Fig.\ref{fig:scaling}(a)), irrespective of the value of $J_K$. This was pointed out in Ref.\onlinecite{Oh2006} and interpreted as a pseudo-orthogonality catastrophe: this is the only site of the chain with correlations large enough to overcome the PPT condition of separability, Eq.(\ref{concurrence}). 

\begin{figure}
 \includegraphics[height=0.48\textwidth, angle=-90]{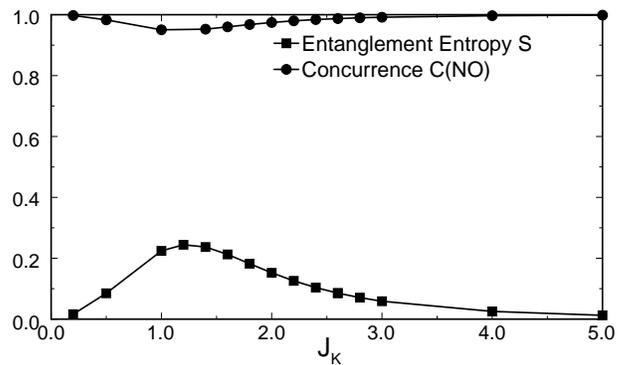}
\caption{Concurrence with the dominant natural orbital, and von Neuman bi-partite entanglement entropy between the ``block'' formed by the impurity and the natural orbital, and the rest of the system.  
} 
\label{fig:entropy}
\end{figure}


\section{Projected natural orbitals}
It would be highly desirable to be able to disentangle the Kondo singlet from the rest of the Fermi sea. Since presumably the impurity is entangled to a {\it single} electron, all we need is to identify this electronic wave function. This can be done by considering  a particular basis of ``projected'' natural orbitals.
We define the ``projected'' single particle Green's function as:
\[
\tilde{G}_{ij}^\sigma = -\langle S^z_{\rm imp} c^\dagger_{\sigma j}c_{\sigma i} \rangle.
\]
The natural orbitals (NOs) are defined as the eigenvectors of this matrix $|\alpha_n\rangle = \alpha^\dagger_n|\rm{vac}\rangle$ (we have omitted the spin index for simplicity). 
Unlike the single particle Green's function, this projection will unambiguously yield a dominant single particle eigenstate (See Appendix B). This wave-function $|\alpha_0\rangle$ has eigenvalue $\sim 1/4$, and all the other eigenvalues are close to zero for the entire range of $J_K$ considered. 
In Fig.\ref{fig:no}(a) and (b) we show the wave function coefficients of the dominant natural orbital decomposed into its real-space components $|\alpha_0\rangle =\sum_i a_i c^\dagger_i|\rm{vac}\rangle$, and momentum basis $|\alpha_0\rangle =\sum_k w_k c^\dagger_k|\rm{vac}\rangle$.
This wave function is equal to $\alpha^\dagger_0=c^\dagger_{k_F}$ for $J_K \rightarrow 0$, and $\alpha^\dagger_0 = c^\dagger_{r_0}$ for $J_K \rightarrow \infty$. 

\begin{figure}
 \includegraphics[height=0.48\textwidth, angle=-90]{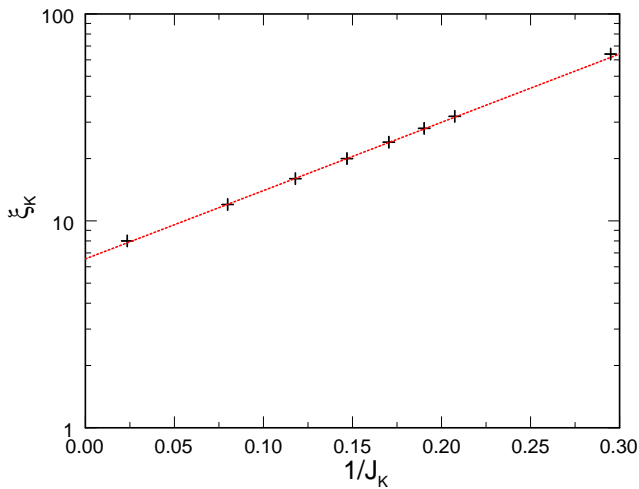}
\caption{Kondo screening length extracted from a finite-size analysis by using the concurrence as an ``order parameter''. The dashed line corresponds to an exponential fit $\xi_K=6.55 \exp{(7.6/J_K)}$.
} 
\label{fig:cloud}
\end{figure}

In addition, when calculating the concurrence with the natural orbitals, we find that more than $90\%$ of the entanglement is with the single particle wave function $|\alpha_0\rangle.$ This is illustrated in Fig.\ref{fig:entropy}, where we show the concurrence with the dominant NO as a function of $J_K$.  
These results lead us to a very simple and compact expression for the ground-state wave function, that is rigorously valid in the $J_K \rightarrow 0$ and $J_K \rightarrow \infty$ limits:
\begin{equation}
|{\rm g.s.}\rangle = \frac{1}{\sqrt{2}}\left(|\Uparrow\rangle \alpha^\dagger_{0\downarrow}-|\Downarrow\rangle \alpha^\dagger_{0\uparrow}\right) |{\rm FS'}\rangle,
\label{gs}
\end{equation}
where the impurity spin forms a singlet with the natural orbital $|\alpha_0\rangle$ in a product state (completely disentangled) from a Fermi sea formed by the remaining orthogonal natural orbitals $|{\rm FS'}\rangle = \prod^{(N-1)/2}_{n=1} \alpha^\dagger_{n\uparrow}\alpha^\dagger_{n\downarrow} |\rm{vac}\rangle$.
This wave function is very similar to the one proposed by Yosida in the 60s\cite{Yosida1966,Varma1976Yafet}, and also to the one proposed by Bergmann in his artificial resonant state approach \cite{Bergmann2007}. 

To verify this assumption we calculate the coefficients $a_n$, $b_n$, $c_n$, and $d_n$ of the ground state, Eq.(\ref{gs2}) in this basis. We find that for the dominant natural orbital, $b_0^2 > 0.49$ and $a^2_0,c^2_0,d^2_0 < 0.01$ for all values of $J_K$ considered. All other natural orbitals are either double occupied or empty. In addition, we calculate the bipartite von Neumann entanglement entropy between a subsystem formed by the impurity and the dominant natural orbital, and the rest of the conduction electrons, $S= -\mathrm{Tr}(\rho \log_2 \rho)$, where $\rho$ is the two-particle reduced density matrix of the subsystem. If Eq.(\ref{gs}) was rigorously correct, this quantity should be zero. Our results in Fig.\ref{fig:entropy} show that $S$ is small, particularly in the weak and strong coupling regimes. Even though there is a residual entanglement with the rest of the electrons, this occurs for intermediate values of $J_K \sim W$ which are relevant to ``Kondo box'' physics. In the thermodynamic limit, the physics flows toward the strong coupling regime and we expect $S \rightarrow 0$. One can readily validate this wave-function as a  very good approximation to the actual ground-state by numerically calculating its variational energy: We find that it yields the correct ground state energy with three and even up to four digits (not shown here). 

\section{Kondo screening length}

It has been proposed that the size of the electron wave function that screens the localized spin $\xi_K$ could be measured in mesoscopic devices\cite{Thimm1999, Affleck2001, Simon2001, simon2003kondo, kaul2006spectroscopy, Pereira2008}. Determining this quantity is a non-trivial task, since the wave function typically decays algebraically. However, one could identify this length with the typical system size at which the renormalization flow enters the strong coupling universal regime. In our case, the concurrence at $k_F$ plays the role of determining precisely this cutoff. 

To justify these arguments, we look at the evolution of the spectrum with $J_K$, shown in Fig.\ref{fig:scaling}(b)-(c). The behavior of the energies as a function of $J_K$ bears a resemblance to the NRG spectrum as a function of system size \cite{Cornaglia2002}. 
One can see levels that run parallel to each other: the state labeled $\Delta_2$ correspond to  a particle-hole excitation in the Fermi sea $|{\rm FS'}\rangle$ in Eq.(\ref{gs}) and the level spacing is determined by the system size.
At small $J_K$, one can see that the low energy excitations $\Delta_1$ are being pushed up in energy. These are genuine excitations of the Kondo singlet, and determine the characteristic energy scale $T_K$. 
The crossover between the two regimes happens as the first excited eigenvalue ``merges'' with the single particle excitation. Even though this is not a sharp transition, there is an energy cutoff that is determined by the critical value $J_K^*(L)$ at which the concurrence vanishes as seen in Figs.~\ref{fig:scaling}(a)-(c) for two system sizes.

\begin{figure}
 \includegraphics[height=0.48\textwidth, angle=-90]{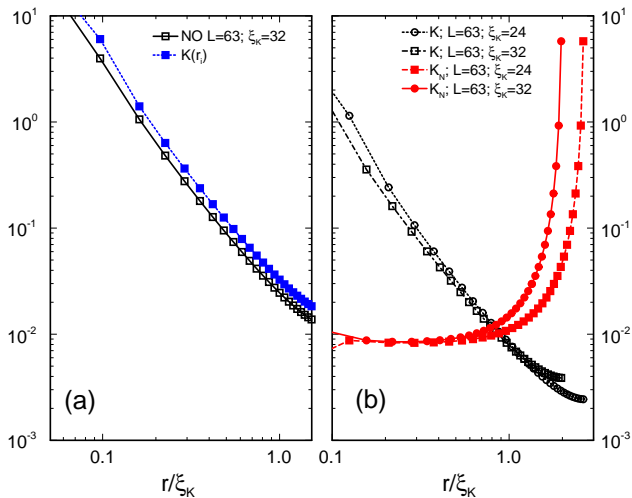}
\caption{(a) Dominant natural orbital and RKKY correlation $K(r_i)$ as defined in the text, for a chain of size $L=63$, and $\xi_K = 32$. The correlation has been multiplied by a factor 4. (b) RKKY correlation with the first and last sites of the chain $K(r_i)$ and $K_N(r_i)$ for two values of $\xi_K$.
}
\label{fig:szsz}
\end{figure}

One can now calculate the critical value $J_K^*(L)$ for different system sizes $L$, and inverting this dependence obtain $\xi_K(J_K)$, shown in Fig.\ref{fig:cloud}. The overall dependence can be summarized very accurately in a fit $\xi_K=6.55\exp{(7.6/J_K)}$. Notice that in the limit $J_K \rightarrow \infty$ the screening length should scale to $\xi_K=1$, implying the possibility of other corrections to the prefactor. For small systems of length $N=5$ the concurrence never vanishes, naturally providing a lattice/energy cutoff for the validity of the model. 
This expression differs from the one obtained using scaling arguments in Refs.\onlinecite{Sorensen1996,Pereira2008}, $\xi_K \sim \exp(\pi/c)$ with $c=2J_K/(1-3J_K^2/4)$ in the strong coupling limit. We believe the discrepancy stems from the fact that we are measuring two different quantities: while theirs corresponds to a scaling length, we are measuring the characteristic ``size'' of the natural orbitals. To illustrate this, we first notice that the natural orbital wavefunction can be directly associated to the amplitude of the RKKY spin-spin correlations\cite{hand2006spin} $K(r_i)=|Z(r_i)-Z(r_{i+1})|$ with $Z(r_i) = \langle S^z_{\rm imp} S^z_i \rangle$, as shown in Fig.\ref{fig:szsz}(a). 
Therefore, by studying the behavior of the correlations, we can now assign a physical meaning to the length $\xi_K$. In Fig.\ref{fig:szsz}(b) we show both the RKKY correlations with the impurity $K(r_i)$, and the ones with the last site of the chain, $K_N=|Z_N(r_i)-Z_N(r_{i+1})|$ with $Z_N(r_i) = \langle S^z_{N} S^z_i \rangle$, for two sets of parameters. The correlations cross at $r=\xi_K$, indicating that beyond this distance from the impurity, the physics is dominated by the free-fermion physics of particles in a box of size $N-\xi_K$ confined between the Kondo cloud and the end of the chain.

\section{Conclusions}
In mesoscopic systems and for small $J_K \sim W/L$, the Kondo cloud does not have space to form and the impurity entangles mainly to the state at the Fermi level.
As $J_K$ increases, the impurity starts creating particle-hole excitations in the Fermi sea and will entangle to the electrons {\it above} the chemical potential, and to the {\it holes} below it. 
Most interestingly, the electrons around the Fermi level tend to decouple from the spin, oblivious to its presence as though the impurity were transparent. This transition from weak to strong coupling regimes is described dramatically by the behavior of the concurrence at the Fermi level.

 Our results for the concurrence with the state at $k_F$ indicate that this measure can be employed as a quasi-order parameter to determine the transition (actually a crossover) from the weak to strong coupling regimes at the point at which it becomes identically zero. This criterion offers the means to measure the Kondo screening length. 

We have identified a dominant single particle wave function that is entangled to the impurity forming a singlet that is, to a great extent, practically disentangled from the rest of the conduction electrons.
We point out that conducting DMRG simulations in the natural orbital basis increases the efficiency in a very dramatic way, even though the Hamiltonian will now involve long range terms\cite{Wolf2014}. This is because the ground state wave function is very close to a product state. For the largest systems studied, keeping of the order of 40 DMRG states may yield machine precision accuracy for the largest systems studied. It is important to notice that these natural orbitals are optimized for the ground state, and not the excited states. 

We have associated the Kondo screening length $\xi_K$ to a characteristic ``size'' of the natural orbital wave-function. This quantity differs from the Kondo scaling length, and corresponds to a property of the wave-function. We have found that beyond this distance from the Kondo impurity the physics is dominated by the boundary, with free fermions weakly correlated to the impurity that ``see'' the Kondo cloud as a scattering center.

These results and the tools developed in this work provide new insight into the single impurity problem, and can lead to a better understanding of heavy fermion systems and exhaustion physics, and efficient real-frequency impurity solvers for dynamical mean field theory calculations\cite{Georges1996,Garcia2004,Zgid2012,Lu2014,Wolf2014b}.
Work in this direction is currently being conducted and presented elsewhere.

\acknowledgments
We thank I. Hamad for discussions that motivated our interest in this problem, and G. Martins, F. Heidrich-Meisner, R. G. Pereira, and E. Katz for useful comments.
The authors acknowledge the U.S. Department of Energy, Office of Basic Energy Sciences, for support under grant DE-SC0014407.

\section*{APPENDIX A: TWO-PARTICLE DENSITY MATRIX}

In this appendix we describe the identities that yield the simple form of the
the two-particle density matrix $\rho$ in the presence of time-reversal symmetry, which is defined as
\[
\rho_{\alpha \beta, \alpha' \beta'} = \frac{1}{2}\langle c^\dagger_{2\beta} c^\dagger_{1\alpha} c_{1\alpha'} c_{2\beta'}\rangle.
\]
Using this expression, one can re-write it more explicitly in a matrix form:
\begin{equation}
\rho = \frac{1}{2}\left(
\begin{array}{cccc}
\langle n_{1\uparrow}n_{2\uparrow}\rangle & 0 & 0 & 0 \\
0 & \langle n_{1\uparrow}n_{2\downarrow}\rangle & \langle S^+_1S^-_2\rangle & 0 \\
0 & \langle S^+_2S^-_1\rangle & \langle n_{1\downarrow}n_{2\uparrow}\rangle & 0 \\
0 & 0 & 0 & \langle n_{1\downarrow}n_{2\downarrow}\rangle
\end{array}
\right).
\end{equation}
Assuming that the site labeled as 1 corresponds to the impurity, we introduce the identities:
\begin{eqnarray}
\langle N_1 n_{2\sigma} \rangle &=& \langle n_{1\uparrow}n_{2\sigma} \rangle + \langle n_{1\downarrow}n_{2\sigma} \rangle = \langle n_{2\sigma}  \rangle,
\end{eqnarray}
which yield:
\begin{eqnarray}
\langle n_{1\uparrow}n_{2\sigma} \rangle &=& \frac{1}{2}\langle n_{2\sigma} \rangle + \langle S^z_1 n_{2\sigma} \rangle \nonumber \\
\langle n_{1\downarrow}n_{2\sigma} \rangle &=& \frac{1}{2}\langle n_{2\sigma} \rangle - \langle S^z_1 n_{2\sigma} \rangle.
\label{eq1}
\end{eqnarray}

In addition,
\begin{eqnarray}
\langle S^z_1 n_{2\uparrow} \rangle &=& \frac{1}{2}\left( \langle n_{1\uparrow}n_{2\uparrow} \rangle - \langle n_{1\downarrow}n_{2\uparrow}\rangle\right) \nonumber \\
\langle S^z_1 n_{2\downarrow} \rangle &=& \frac{1}{2}\left( \langle n_{1\uparrow}n_{2\downarrow} \rangle - \langle n_{1\downarrow}n_{2\downarrow}\rangle\right).\end{eqnarray}
Since $\langle n_{1\uparrow}n_{2\uparrow} \rangle = \langle n_{1\downarrow}n_{2\downarrow}\rangle$ and $\langle n_{1\downarrow}n_{2\uparrow}\rangle=\langle n_{1\uparrow}n_{2\downarrow} \rangle$, we obtain that
\[
\langle S^z_1 n_{2\uparrow} \rangle = -\langle S^z_1 n_{2\downarrow} \rangle.
\]
Replacing in Eq.(\ref{eq1}):
\begin{eqnarray}
\langle n_{1\uparrow}n_{2\sigma} \rangle &=& \frac{1}{2}\langle n_{2\uparrow} \rangle +\sigma \langle S^z_1 n_{2\uparrow} \rangle \nonumber \\
\langle n_{1\downarrow}n_{2\sigma} \rangle &=& \frac{1}{2}\langle n_{2\uparrow} \rangle -\sigma \langle S^z_1 n_{2\uparrow} \rangle.
\end{eqnarray}
where we have used $\langle n_{2\uparrow} \rangle = \langle n_{2\downarrow} \rangle$, and $\sigma = \pm$ for up, and down spins.
We resort to the $SU(2)$ symmetry of the problem to find:
\[
\langle S^+_1 S^-_2 \rangle = \langle S^+_1 S^-_2 \rangle = 2 \langle S^z_1 S^z_2 \rangle = 2\langle S^z_1 n_{2\uparrow} \rangle.
\]
Defining $\delta = -2\langle S^z_1 n_{2\uparrow} \rangle$, we finally obtain
\begin{equation}
\rho = \frac{1}{2}\left(
\begin{array}{cccc}
\frac{\langle n_{2\uparrow}\rangle-\delta}{2} & 0 & 0 & 0 \\
0 & \frac{\langle n_{2\uparrow}\rangle+\delta}{2} & -\delta & 0 \\
0 & -\delta & \frac{\langle n_{2\uparrow}\rangle+\delta}{2} & 0 \\
0 & 0 & 0 & \frac{\langle n_{2\uparrow}\rangle-\delta }{2}
\end{array}
\right),
\end{equation}
which is the desired result, as described in the text.

\section*{APPENDIX B: NATURAL ORBITALS}

Let us assume that the ground state can be written as
\begin{equation}
|{\rm g.s.}\rangle = \frac{1}{\sqrt{2}}\left[|\Uparrow\rangle \alpha^\dagger_{0\downarrow}-|\Downarrow\rangle \alpha^\dagger_{0\uparrow}\right] \prod_{n=1}^{N/2} \alpha^\dagger_{n\uparrow}\alpha^\dagger_{n\downarrow} |\rm{vac}\rangle,
\label{gs3}
\end{equation}
where the single particle states $|\alpha_i\rangle$ are the natural orbitals. Without loss of generality, we can calculate the single particle Green's function in this basis $G^\uparrow_{ij}=\langle \alpha^\dagger_{i\uparrow} \alpha_{j\uparrow}\rangle$, which is diagonal with elements $G^\uparrow_{00}=1/2$, $G^\uparrow_{ii}=1$ for $0<i\le N$. Notice that this corresponds precisely to the momentum distribution function $n(k)$ in the weak coupling limit, as discussed in the text.

This means that their weights are pretty much evenly distributed, without assigning a dominant weight to a single state. For this reason, we introduce the ``projected'' single particle Green's function as $\tilde{G}^\uparrow_{ij}= -\langle S^z_{\rm imp} \alpha^\dagger_{i\uparrow} \alpha_{j\uparrow}\rangle$, which is also diagonal, but with eigenvalues $\tilde{G}^\uparrow_{00}=1/4$ and $\tilde{G}^\uparrow_{ii}=0$ for $0<i\le N/2$, clearly identifying the single particle state that is entangled to the impurity. Even though both formulations yield the same eigenfunctions, in a generic scenario, where the ground state is approximately, but not exactly, described by Eq.(\ref{gs}), the projected Green's function can always identify the dominant orbital with high level of accuracy, as shown in the main text, while the conventional formulation does not.


\begin{thebibliography}{64}
\expandafter\ifx\csname natexlab\endcsname\relax\def\natexlab#1{#1}\fi
\expandafter\ifx\csname bibnamefont\endcsname\relax
  \def\bibnamefont#1{#1}\fi
\expandafter\ifx\csname bibfnamefont\endcsname\relax
  \def\bibfnamefont#1{#1}\fi
\expandafter\ifx\csname citenamefont\endcsname\relax
  \def\citenamefont#1{#1}\fi
\expandafter\ifx\csname url\endcsname\relax
  \def\url#1{\texttt{#1}}\fi
\expandafter\ifx\csname urlprefix\endcsname\relax\def\urlprefix{URL }\fi
\providecommand{\bibinfo}[2]{#2}
\providecommand{\eprint}[2][]{\url{#2}}

\bibitem[{\citenamefont{Hewson}(1997)}]{HewsonBook}
\bibinfo{author}{\bibfnamefont{A.}~\bibnamefont{Hewson}},
  \emph{\bibinfo{title}{The Kondo Problem to Heavy Fermions}}
  (\bibinfo{publisher}{Cambridge Univ. Press}, \bibinfo{year}{1997}).

\bibitem[{\citenamefont{Cox and Zawadowski}(1998)}]{cox1998exotic}
\bibinfo{author}{\bibfnamefont{D.}~\bibnamefont{Cox}} \bibnamefont{and}
  \bibinfo{author}{\bibfnamefont{A.}~\bibnamefont{Zawadowski}},
  \bibinfo{journal}{Advances in Physics} \textbf{\bibinfo{volume}{47}},
  \bibinfo{pages}{599} (\bibinfo{year}{1998}).

\bibitem[{\citenamefont{Wilson}(1975)}]{nrg_wilson}
\bibinfo{author}{\bibfnamefont{K.~G.} \bibnamefont{Wilson}},
  \bibinfo{journal}{Rev. Mod. Phys.} \textbf{\bibinfo{volume}{47}},
  \bibinfo{pages}{773} (\bibinfo{year}{1975}).

\bibitem[{\citenamefont{Bulla et~al.}(2008)\citenamefont{Bulla, Costi, and
  Pruschke}}]{nrg_bulla}
\bibinfo{author}{\bibfnamefont{R.}~\bibnamefont{Bulla}},
  \bibinfo{author}{\bibfnamefont{T.~A.} \bibnamefont{Costi}}, \bibnamefont{and}
  \bibinfo{author}{\bibfnamefont{T.}~\bibnamefont{Pruschke}},
  \bibinfo{journal}{Rev. Mod. Phys.} \textbf{\bibinfo{volume}{80}},
  \bibinfo{pages}{395} (\bibinfo{year}{2008}).

\bibitem[{\citenamefont{Andrei et~al.}(1983)\citenamefont{Andrei, Furuya, and
  Lowenstein}}]{andrei1983solution}
\bibinfo{author}{\bibfnamefont{N.}~\bibnamefont{Andrei}},
  \bibinfo{author}{\bibfnamefont{K.}~\bibnamefont{Furuya}}, \bibnamefont{and}
  \bibinfo{author}{\bibfnamefont{J.}~\bibnamefont{Lowenstein}},
  \bibinfo{journal}{Rev. Mod. Phys.} \textbf{\bibinfo{volume}{55}},
  \bibinfo{pages}{331} (\bibinfo{year}{1983}).

\bibitem[{\citenamefont{Tsvelik and Wiegmann}(1983)}]{tsvelik1983I}
\bibinfo{author}{\bibfnamefont{A.}~\bibnamefont{Tsvelik}} \bibnamefont{and}
  \bibinfo{author}{\bibfnamefont{P.}~\bibnamefont{Wiegmann}},
  \bibinfo{journal}{J. Phys. C:Solid State Phys.}
  \textbf{\bibinfo{volume}{16}}, \bibinfo{pages}{2281} (\bibinfo{year}{1983}).

\bibitem[{\citenamefont{B\"usser et~al.}(2013)\citenamefont{B\"usser, Martins,
  and Feiguin}}]{busser2013lanczos}
\bibinfo{author}{\bibfnamefont{C.~A.} \bibnamefont{B\"usser}},
  \bibinfo{author}{\bibfnamefont{G.~B.} \bibnamefont{Martins}},
  \bibnamefont{and} \bibinfo{author}{\bibfnamefont{A.~E.}
  \bibnamefont{Feiguin}}, \bibinfo{journal}{Phys. Rev. B}
  \textbf{\bibinfo{volume}{88}}, \bibinfo{pages}{245113}
  (\bibinfo{year}{2013}).

\bibitem[{\citenamefont{Anderson}(1970)}]{Anderson1970}
\bibinfo{author}{\bibfnamefont{P.~W.} \bibnamefont{Anderson}},
  \bibinfo{journal}{J. Phys. C: Solid State Phys.}
  \textbf{\bibinfo{volume}{3}}, \bibinfo{pages}{2436} (\bibinfo{year}{1970}).

\bibitem[{\citenamefont{Sorensen and Affleck}(1996)}]{Sorensen1996}
\bibinfo{author}{\bibnamefont{Sorensen}} \bibnamefont{and}
  \bibinfo{author}{\bibnamefont{Affleck}}, \bibinfo{journal}{Phys. Rev. B}
  \textbf{\bibinfo{volume}{53}}, \bibinfo{pages}{9153} (\bibinfo{year}{1996}).

\bibitem[{\citenamefont{Affleck and
  Simon}(2001{\natexlab{a}})}]{affleck2001detecting}
\bibinfo{author}{\bibfnamefont{I.}~\bibnamefont{Affleck}} \bibnamefont{and}
  \bibinfo{author}{\bibfnamefont{P.}~\bibnamefont{Simon}},
  \bibinfo{journal}{Phys. Rev. Lett.} \textbf{\bibinfo{volume}{86}},
  \bibinfo{pages}{2854} (\bibinfo{year}{2001}{\natexlab{a}}).

\bibitem[{\citenamefont{Sorensen and Affleck}(2005)}]{Sorensen2005a}
\bibinfo{author}{\bibfnamefont{E.~S.} \bibnamefont{Sorensen}} \bibnamefont{and}
  \bibinfo{author}{\bibfnamefont{I.}~\bibnamefont{Affleck}},
  \bibinfo{journal}{Phys. Rev. Lett.} \textbf{\bibinfo{volume}{94}},
  \bibinfo{pages}{086601} (\bibinfo{year}{2005}).

\bibitem[{\citenamefont{Affleck}(2010)}]{affleck2009kondo}
\bibinfo{author}{\bibfnamefont{I.}~\bibnamefont{Affleck}}, in
  \emph{\bibinfo{booktitle}{Perspectives on Mesoscopic Physics: Dedicated to
  Professor Yoseph Imry's 70th Birthday}}, edited by
  \bibinfo{editor}{\bibfnamefont{A.}~\bibnamefont{Aharony}} \bibnamefont{and}
  \bibinfo{editor}{\bibfnamefont{O.}~\bibnamefont{Entin-Wohlman}}
  (\bibinfo{publisher}{World Scientific, Singapore}, \bibinfo{year}{2010}), pp.
  \bibinfo{pages}{1--44}.

\bibitem[{\citenamefont{Bergmann}(2008)}]{bergmann08}
\bibinfo{author}{\bibfnamefont{G.}~\bibnamefont{Bergmann}},
  \bibinfo{journal}{Phys. Rev. B} \textbf{\bibinfo{volume}{77}},
  \bibinfo{pages}{104401} (\bibinfo{year}{2008}).

\bibitem[{\citenamefont{Holzner et~al.}(2009)\citenamefont{Holzner, McCulloch,
  Schollw\"ock, von Delft, and Heidrich-Meisner}}]{FHM09}
\bibinfo{author}{\bibfnamefont{A.}~\bibnamefont{Holzner}},
  \bibinfo{author}{\bibfnamefont{I.~P.} \bibnamefont{McCulloch}},
  \bibinfo{author}{\bibfnamefont{U.}~\bibnamefont{Schollw\"ock}},
  \bibinfo{author}{\bibfnamefont{J.}~\bibnamefont{von Delft}},
  \bibnamefont{and}
  \bibinfo{author}{\bibfnamefont{F.}~\bibnamefont{Heidrich-Meisner}},
  \bibinfo{journal}{Phys. Rev. B} \textbf{\bibinfo{volume}{80}},
  \bibinfo{pages}{205114} (\bibinfo{year}{2009}).

\bibitem[{\citenamefont{B\"usser et~al.}(2010)\citenamefont{B\"usser, Martins,
  Ribeiro, E.~Vernek, and Dagotto}}]{Busser10}
\bibinfo{author}{\bibfnamefont{C.~A.} \bibnamefont{B\"usser}},
  \bibinfo{author}{\bibfnamefont{G.~B.} \bibnamefont{Martins}},
  \bibinfo{author}{\bibfnamefont{L.} \bibnamefont{Costa Ribeiro}},
  \bibinfo{author}{\bibnamefont{E.~Vernek}},
  \bibinfo{author}{\bibfnamefont{E.~V.~Anda}} ,
  \bibnamefont{and} \bibinfo{author}{\bibfnamefont{E.}~\bibnamefont{Dagotto}},
  \bibinfo{journal}{Phys. Rev. B} \textbf{\bibinfo{volume}{81}},
  \bibinfo{pages}{045111} (\bibinfo{year}{2010}).

\bibitem[{\citenamefont{Schlottmann}(2001)}]{schlottmann2001kondo}
\bibinfo{author}{\bibfnamefont{P.}~\bibnamefont{Schlottmann}},
  \bibinfo{journal}{Phys. Rev. B} \textbf{\bibinfo{volume}{65}},
  \bibinfo{pages}{024420} (\bibinfo{year}{2001}).

\bibitem[{\citenamefont{Simon and Affleck}(2002)}]{simon2002finite}
\bibinfo{author}{\bibfnamefont{P.}~\bibnamefont{Simon}} \bibnamefont{and}
  \bibinfo{author}{\bibfnamefont{I.}~\bibnamefont{Affleck}},
  \bibinfo{journal}{Phys. Rev. Lett.} \textbf{\bibinfo{volume}{89}},
  \bibinfo{pages}{206602} (\bibinfo{year}{2002}).

\bibitem[{\citenamefont{Simon and Affleck}(2003)}]{simon2003kondo}
\bibinfo{author}{\bibfnamefont{P.}~\bibnamefont{Simon}} \bibnamefont{and}
  \bibinfo{author}{\bibfnamefont{I.}~\bibnamefont{Affleck}},
  \bibinfo{journal}{Phys. Rev. B} \textbf{\bibinfo{volume}{68}},
  \bibinfo{pages}{115304} (\bibinfo{year}{2003}).

\bibitem[{\citenamefont{Kaul et~al.}(2006)\citenamefont{Kaul, Zar{\'a}nd,
  Chandrasekharan, Ullmo, and Baranger}}]{kaul2006spectroscopy}
\bibinfo{author}{\bibfnamefont{R.~K.} \bibnamefont{Kaul}},
  \bibinfo{author}{\bibfnamefont{G.}~\bibnamefont{Zar{\'a}nd}},
  \bibinfo{author}{\bibfnamefont{S.}~\bibnamefont{Chandrasekharan}},
  \bibinfo{author}{\bibfnamefont{D.}~\bibnamefont{Ullmo}}, \bibnamefont{and}
  \bibinfo{author}{\bibfnamefont{H.~U.} \bibnamefont{Baranger}},
  \bibinfo{journal}{Phys. Rev. Lett.} \textbf{\bibinfo{volume}{96}},
  \bibinfo{pages}{176802} (\bibinfo{year}{2006}).

\bibitem[{\citenamefont{Hand et~al.}(2006)\citenamefont{Hand, Kroha, and
  Monien}}]{hand2006spin}
\bibinfo{author}{\bibfnamefont{T.}~\bibnamefont{Hand}},
  \bibinfo{author}{\bibfnamefont{J.}~\bibnamefont{Kroha}}, \bibnamefont{and}
  \bibinfo{author}{\bibfnamefont{H.}~\bibnamefont{Monien}},
  \bibinfo{journal}{Phys. Rev. Lett.} \textbf{\bibinfo{volume}{97}},
  \bibinfo{pages}{136604} (\bibinfo{year}{2006}).

\bibitem[{\citenamefont{{Nozi\`eres, P.}}(1976)}]{Nozieres1976}
\bibinfo{author}{\bibnamefont{{Nozi\`eres, P.}}}, \bibinfo{journal}{J. Phys.
  Colloques} \textbf{\bibinfo{volume}{37}}, \bibinfo{pages}{C1}
  (\bibinfo{year}{1976}).

\bibitem[{\citenamefont{Nozi{\`e}res and Blandin}(1980)}]{Nozieres1980}
\bibinfo{author}{\bibfnamefont{P.}~\bibnamefont{Nozi{\`e}res}}
  \bibnamefont{and} \bibinfo{author}{\bibfnamefont{A.}~\bibnamefont{Blandin}},
  \bibinfo{journal}{{Journal de Physique}} \textbf{\bibinfo{volume}{41}},
  \bibinfo{pages}{193} (\bibinfo{year}{1980}).

\bibitem[{\citenamefont{Coleman}(2002)}]{Coleman2002}
\bibinfo{author}{\bibfnamefont{P.}~\bibnamefont{Coleman}},
  \bibinfo{journal}{AIP Conference Proceedings} \textbf{\bibinfo{volume}{629}},
  \bibinfo{pages}{79} (\bibinfo{year}{2002}).

\bibitem[{\citenamefont{{Nozi\`eres, P.}}(1985)}]{Nozieres1985}
\bibinfo{author}{\bibnamefont{{Nozi\`eres, P.}}}, \bibinfo{journal}{Ann. Phys.
  Fr.} \textbf{\bibinfo{volume}{10}}, \bibinfo{pages}{19}
  (\bibinfo{year}{1985}).

\bibitem[{\citenamefont{Nozi\`eres}(1998)}]{Nozieres1998}
\bibinfo{author}{\bibfnamefont{P.}~\bibnamefont{Nozi\`eres}},
  \bibinfo{journal}{Eur. Phys. J. B} \textbf{\bibinfo{volume}{6}},
  \bibinfo{pages}{447–457} (\bibinfo{year}{1998}).

\bibitem[{\citenamefont{Coleman et~al.}(2001)\citenamefont{Coleman, P\'epin,
  Si, and Ramazashvili}}]{Coleman2001}
\bibinfo{author}{\bibfnamefont{P.}~\bibnamefont{Coleman}},
  \bibinfo{author}{\bibfnamefont{C.}~\bibnamefont{P\'epin}},
  \bibinfo{author}{\bibfnamefont{Q.}~\bibnamefont{Si}}, \bibnamefont{and}
  \bibinfo{author}{\bibfnamefont{R.}~\bibnamefont{Ramazashvili}},
  \bibinfo{journal}{Journal of Physics: Condensed Matter}
  \textbf{\bibinfo{volume}{13}}, \bibinfo{pages}{R723} (\bibinfo{year}{2001}).

\bibitem[{\citenamefont{Hackl and Vojta}(2008)}]{Hackl2008}
\bibinfo{author}{\bibfnamefont{A.}~\bibnamefont{Hackl}} \bibnamefont{and}
  \bibinfo{author}{\bibfnamefont{M.}~\bibnamefont{Vojta}},
  \bibinfo{journal}{Phys. Rev. B} \textbf{\bibinfo{volume}{77}},
  \bibinfo{pages}{134439} (\bibinfo{year}{2008}), \eprint{0712.2107}.

\bibitem[{\citenamefont{Feiguin and B{\"u}sser}(2011)}]{feiguin2011reducing}
\bibinfo{author}{\bibfnamefont{A.~E.} \bibnamefont{Feiguin}} \bibnamefont{and}
  \bibinfo{author}{\bibfnamefont{C.~A.} \bibnamefont{B{\"u}sser}},
  \bibinfo{journal}{Phys. Rev. B} \textbf{\bibinfo{volume}{84}},
  \bibinfo{pages}{115403} (\bibinfo{year}{2011}).

\bibitem[{\citenamefont{B{\"u}sser and Feiguin}(2012)}]{busser2012designing}
\bibinfo{author}{\bibfnamefont{C.~A.} \bibnamefont{B{\"u}sser}}
  \bibnamefont{and} \bibinfo{author}{\bibfnamefont{A.~E.}
  \bibnamefont{Feiguin}}, \bibinfo{journal}{Phys. Rev. B}
  \textbf{\bibinfo{volume}{86}}, \bibinfo{pages}{165410}
  (\bibinfo{year}{2012}).

\bibitem[{\citenamefont{White}(1992)}]{White1992}
\bibinfo{author}{\bibfnamefont{S.~R.} \bibnamefont{White}},
  \bibinfo{journal}{Phys. Rev. Lett.} \textbf{\bibinfo{volume}{69}},
  \bibinfo{pages}{2863} (\bibinfo{year}{1992}).

\bibitem[{\citenamefont{White}(1993)}]{White1993}
\bibinfo{author}{\bibfnamefont{S.~R.} \bibnamefont{White}},
  \bibinfo{journal}{Phys. Rev. B} \textbf{\bibinfo{volume}{48}},
  \bibinfo{pages}{10345} (\bibinfo{year}{1993}).

\bibitem[{\citenamefont{Schollw{\"o}ck}(2005)}]{Schollwock2005density}
\bibinfo{author}{\bibfnamefont{U.}~\bibnamefont{Schollw{\"o}ck}},
  \bibinfo{journal}{Rev. Mod. Phys.} \textbf{\bibinfo{volume}{77}},
  \bibinfo{pages}{259} (\bibinfo{year}{2005}).

\bibitem[{\citenamefont{S{\o}rensen et~al.}(2007)\citenamefont{S{\o}rensen,
  Chang, Laflorencie, and Affleck}}]{sorensen2007}
\bibinfo{author}{\bibfnamefont{E.~S.} \bibnamefont{S{\o}rensen}},
  \bibinfo{author}{\bibfnamefont{M.-S.} \bibnamefont{Chang}},
  \bibinfo{author}{\bibfnamefont{N.}~\bibnamefont{Laflorencie}},
  \bibnamefont{and} \bibinfo{author}{\bibfnamefont{I.}~\bibnamefont{Affleck}},
  \bibinfo{journal}{J. Stat. Mech.} \textbf{\bibinfo{volume}{2007}},
  \bibinfo{pages}{L01001} (\bibinfo{year}{2007}).

\bibitem[{\citenamefont{Affleck et~al.}(2009)\citenamefont{Affleck,
  Laflorencie, and S{\o}rensen}}]{affleck2009}
\bibinfo{author}{\bibfnamefont{I.}~\bibnamefont{Affleck}},
  \bibinfo{author}{\bibfnamefont{N.}~\bibnamefont{Laflorencie}},
  \bibnamefont{and} \bibinfo{author}{\bibfnamefont{E.~S.}
  \bibnamefont{S{\o}rensen}}, \bibinfo{journal}{J. Phys. A: Math. Theor.}
  \textbf{\bibinfo{volume}{42}}, \bibinfo{pages}{504009}
  (\bibinfo{year}{2009}).

\bibitem[{\citenamefont{Cho and McKenzie}(2006)}]{Cho2006}
\bibinfo{author}{\bibfnamefont{S.~Y.} \bibnamefont{Cho}} \bibnamefont{and}
  \bibinfo{author}{\bibfnamefont{R.~H.} \bibnamefont{McKenzie}},
  \bibinfo{journal}{Phys. Rev. A} \textbf{\bibinfo{volume}{73}},
  \bibinfo{pages}{012109} (\bibinfo{year}{2006}).

\bibitem[{\citenamefont{Oh and Kim}(2006)}]{Oh2006}
\bibinfo{author}{\bibfnamefont{S.}~\bibnamefont{Oh}} \bibnamefont{and}
  \bibinfo{author}{\bibfnamefont{J.}~\bibnamefont{Kim}},
  \bibinfo{journal}{Phys. Rev. B} \textbf{\bibinfo{volume}{73}},
  \bibinfo{pages}{052407} (\bibinfo{year}{2006}).

\bibitem[{\citenamefont{Amico et~al.}(2008)\citenamefont{Amico, Fazio,
  Osterloh, and Vedral}}]{Amico2008}
\bibinfo{author}{\bibfnamefont{L.}~\bibnamefont{Amico}},
  \bibinfo{author}{\bibfnamefont{R.}~\bibnamefont{Fazio}},
  \bibinfo{author}{\bibfnamefont{A.}~\bibnamefont{Osterloh}}, \bibnamefont{and}
  \bibinfo{author}{\bibfnamefont{V.}~\bibnamefont{Vedral}},
  \bibinfo{journal}{Rev. Mod. Phys.} \textbf{\bibinfo{volume}{80}},
  \bibinfo{pages}{517} (\bibinfo{year}{2008}).

\bibitem[{\citenamefont{Horodecki et~al.}(2009)\citenamefont{Horodecki,
  Horodecki, Horodecki, and Horodecki}}]{horodecki2009}
\bibinfo{author}{\bibfnamefont{R.}~\bibnamefont{Horodecki}},
  \bibinfo{author}{\bibfnamefont{P.}~\bibnamefont{Horodecki}},
  \bibinfo{author}{\bibfnamefont{M.}~\bibnamefont{Horodecki}},
  \bibnamefont{and}
  \bibinfo{author}{\bibfnamefont{K.}~\bibnamefont{Horodecki}},
  \bibinfo{journal}{Rev. Mod. Phys.} \textbf{\bibinfo{volume}{81}},
  \bibinfo{pages}{865} (\bibinfo{year}{2009}).

\bibitem[{\citenamefont{Vedral et~al.}(1997)\citenamefont{Vedral, Plenio,
  Rippin, and Knight}}]{Vedral1997}
\bibinfo{author}{\bibfnamefont{V.}~\bibnamefont{Vedral}},
  \bibinfo{author}{\bibfnamefont{M.~B.} \bibnamefont{Plenio}},
  \bibinfo{author}{\bibfnamefont{M.~A.} \bibnamefont{Rippin}},
  \bibnamefont{and} \bibinfo{author}{\bibfnamefont{P.~L.}
  \bibnamefont{Knight}}, \bibinfo{journal}{Phys. Rev. Lett.}
  \textbf{\bibinfo{volume}{78}}, \bibinfo{pages}{2275} (\bibinfo{year}{1997}).

\bibitem[{\citenamefont{Hill and Wootters}(1997)}]{Hill1997}
\bibinfo{author}{\bibfnamefont{S.}~\bibnamefont{Hill}} \bibnamefont{and}
  \bibinfo{author}{\bibfnamefont{W.~K.} \bibnamefont{Wootters}},
  \bibinfo{journal}{Phys. Rev. Lett.} \textbf{\bibinfo{volume}{78}},
  \bibinfo{pages}{5022} (\bibinfo{year}{1997}).

\bibitem[{\citenamefont{Wootters}(1998)}]{Wootters1998}
\bibinfo{author}{\bibfnamefont{W.~K.} \bibnamefont{Wootters}},
  \bibinfo{journal}{Phys. Rev. Lett.} \textbf{\bibinfo{volume}{80}},
  \bibinfo{pages}{2245} (\bibinfo{year}{1998}).

\bibitem[{\citenamefont{Vedral and Plenio}(1998)}]{Vedral1998}
\bibinfo{author}{\bibfnamefont{V.}~\bibnamefont{Vedral}} \bibnamefont{and}
  \bibinfo{author}{\bibfnamefont{M.~B.} \bibnamefont{Plenio}},
  \bibinfo{journal}{Phys. Rev. A} \textbf{\bibinfo{volume}{57}},
  \bibinfo{pages}{1619} (\bibinfo{year}{1998}).

\bibitem[{\citenamefont{Verstraete et~al.}(2001)\citenamefont{Verstraete,
  Dehaene, and DeMoor}}]{Verstraete2001}
\bibinfo{author}{\bibfnamefont{F.}~\bibnamefont{Verstraete}},
  \bibinfo{author}{\bibfnamefont{J.}~\bibnamefont{Dehaene}}, \bibnamefont{and}
  \bibinfo{author}{\bibfnamefont{B.}~\bibnamefont{DeMoor}},
  \bibinfo{journal}{Phys. Rev. A} \textbf{\bibinfo{volume}{64}},
  \bibinfo{pages}{010101} (\bibinfo{year}{2001}).

\bibitem[{\citenamefont{Verstraete et~al.}(2003)\citenamefont{Verstraete,
  Dehaene, and De~Moor}}]{Verstraete2003}
\bibinfo{author}{\bibfnamefont{F.}~\bibnamefont{Verstraete}},
  \bibinfo{author}{\bibfnamefont{J.}~\bibnamefont{Dehaene}}, \bibnamefont{and}
  \bibinfo{author}{\bibfnamefont{B.}~\bibnamefont{De~Moor}},
  \bibinfo{journal}{Phys. Rev. A} \textbf{\bibinfo{volume}{68}},
  \bibinfo{pages}{012103} (\bibinfo{year}{2003}).

\bibitem[{\citenamefont{Modi et~al.}(2010)\citenamefont{Modi, Paterek, Son,
  Vedral, and Williamson}}]{Modi2010}
\bibinfo{author}{\bibfnamefont{K.}~\bibnamefont{Modi}},
  \bibinfo{author}{\bibfnamefont{T.}~\bibnamefont{Paterek}},
  \bibinfo{author}{\bibfnamefont{W.}~\bibnamefont{Son}},
  \bibinfo{author}{\bibfnamefont{V.}~\bibnamefont{Vedral}}, \bibnamefont{and}
  \bibinfo{author}{\bibfnamefont{M.}~\bibnamefont{Williamson}},
  \bibinfo{journal}{Phys. Rev. Lett.} \textbf{\bibinfo{volume}{104}},
  \bibinfo{pages}{080501} (\bibinfo{year}{2010}).

\bibitem[{\citenamefont{Bennett et~al.}(1996)\citenamefont{Bennett, DiVincenzo,
  Smolin, and Wootters}}]{Bennett1996}
\bibinfo{author}{\bibfnamefont{C.~H.} \bibnamefont{Bennett}},
  \bibinfo{author}{\bibfnamefont{D.~P.} \bibnamefont{DiVincenzo}},
  \bibinfo{author}{\bibfnamefont{J.~A.} \bibnamefont{Smolin}},
  \bibnamefont{and} \bibinfo{author}{\bibfnamefont{W.~K.}
  \bibnamefont{Wootters}}, \bibinfo{journal}{Phys. Rev. A}
  \textbf{\bibinfo{volume}{54}}, \bibinfo{pages}{3824} (\bibinfo{year}{1996}).

\bibitem[{\citenamefont{Gharibian}(2010)}]{Gharibian2010}
\bibinfo{author}{\bibfnamefont{S.}~\bibnamefont{Gharibian}},
  \bibinfo{journal}{Quantum Information {\&} Computation}
  \textbf{\bibinfo{volume}{10}}, \bibinfo{pages}{343} (\bibinfo{year}{2010}).

\bibitem[{\citenamefont{Peres}(1996)}]{Peres1996}
\bibinfo{author}{\bibfnamefont{A.}~\bibnamefont{Peres}},
  \bibinfo{journal}{Phys. Rev. A} \textbf{\bibinfo{volume}{54}},
  \bibinfo{pages}{2685} (\bibinfo{year}{1996}).

\bibitem[{\citenamefont{Wootters}(2001)}]{Wootters2001}
\bibinfo{author}{\bibfnamefont{W.~K.} \bibnamefont{Wootters}},
  \bibinfo{journal}{Quantum Inf. Comput.} \textbf{\bibinfo{volume}{1}},
  \bibinfo{pages}{27} (\bibinfo{year}{2001}).

\bibitem[{\citenamefont{Werner}(1989)}]{Werner1989}
\bibinfo{author}{\bibfnamefont{R.~F.} \bibnamefont{Werner}},
  \bibinfo{journal}{Phys. Rev. A} \textbf{\bibinfo{volume}{40}},
  \bibinfo{pages}{4277} (\bibinfo{year}{1989}).

\bibitem[{\citenamefont{Yosida}(1966)}]{Yosida1966}
\bibinfo{author}{\bibfnamefont{K.}~\bibnamefont{Yosida}},
  \bibinfo{journal}{Phys. Rev.} \textbf{\bibinfo{volume}{147}},
  \bibinfo{pages}{223} (\bibinfo{year}{1966}).

\bibitem[{\citenamefont{Varma and Yafet}(1976)}]{Varma1976Yafet}
\bibinfo{author}{\bibfnamefont{C.~M.} \bibnamefont{Varma}} \bibnamefont{and}
  \bibinfo{author}{\bibfnamefont{Y.}~\bibnamefont{Yafet}},
  \bibinfo{journal}{Phys. Rev. B} \textbf{\bibinfo{volume}{13}},
  \bibinfo{pages}{2950} (\bibinfo{year}{1976}).

\bibitem[{\citenamefont{Bergmann and Zhang}(2007)}]{Bergmann2007}
\bibinfo{author}{\bibfnamefont{G.}~\bibnamefont{Bergmann}} \bibnamefont{and}
  \bibinfo{author}{\bibfnamefont{L.}~\bibnamefont{Zhang}},
  \bibinfo{journal}{Phys. Rev. B} \textbf{\bibinfo{volume}{76}},
  \bibinfo{pages}{064401} (\bibinfo{year}{2007}).

\bibitem[{\citenamefont{Thimm et~al.}(1999)\citenamefont{Thimm, Kroha, and von
  Delft}}]{Thimm1999}
\bibinfo{author}{\bibfnamefont{W.~B.} \bibnamefont{Thimm}},
  \bibinfo{author}{\bibfnamefont{J.}~\bibnamefont{Kroha}}, \bibnamefont{and}
  \bibinfo{author}{\bibfnamefont{J.}~\bibnamefont{von Delft}},
  \bibinfo{journal}{Phys. Rev. Lett.} \textbf{\bibinfo{volume}{82}},
  \bibinfo{pages}{2143} (\bibinfo{year}{1999}).

\bibitem[{\citenamefont{Affleck and Simon}(2001{\natexlab{b}})}]{Affleck2001}
\bibinfo{author}{\bibfnamefont{I.}~\bibnamefont{Affleck}} \bibnamefont{and}
  \bibinfo{author}{\bibfnamefont{P.}~\bibnamefont{Simon}},
  \bibinfo{journal}{Phys. Rev. Lett.} \textbf{\bibinfo{volume}{86}},
  \bibinfo{pages}{2854} (\bibinfo{year}{2001}{\natexlab{b}}).

\bibitem[{\citenamefont{Simon and Affleck}(2001)}]{Simon2001}
\bibinfo{author}{\bibfnamefont{P.}~\bibnamefont{Simon}} \bibnamefont{and}
  \bibinfo{author}{\bibfnamefont{I.}~\bibnamefont{Affleck}},
  \bibinfo{journal}{Phys. Rev. B} \textbf{\bibinfo{volume}{64}},
  \bibinfo{pages}{085308} (\bibinfo{year}{2001}).

\bibitem[{\citenamefont{Pereira et~al.}(2008)\citenamefont{Pereira,
  Laflorencie, Affleck, and Halperin}}]{Pereira2008}
\bibinfo{author}{\bibfnamefont{R.~G.} \bibnamefont{Pereira}},
  \bibinfo{author}{\bibfnamefont{N.}~\bibnamefont{Laflorencie}},
  \bibinfo{author}{\bibfnamefont{I.}~\bibnamefont{Affleck}}, \bibnamefont{and}
  \bibinfo{author}{\bibfnamefont{B.~I.} \bibnamefont{Halperin}},
  \bibinfo{journal}{Phys. Rev. B} \textbf{\bibinfo{volume}{77}},
  \bibinfo{pages}{125327} (\bibinfo{year}{2008}).

\bibitem[{\citenamefont{Cornaglia and Balseiro}(2002)}]{Cornaglia2002}
\bibinfo{author}{\bibfnamefont{P.~S.} \bibnamefont{Cornaglia}}
  \bibnamefont{and} \bibinfo{author}{\bibfnamefont{C.~A.}
  \bibnamefont{Balseiro}}, \bibinfo{journal}{Phys. Rev. B}
  \textbf{\bibinfo{volume}{66}}, \bibinfo{pages}{115303}
  (\bibinfo{year}{2002}).

\bibitem[{\citenamefont{Wolf et~al.}(2014{\natexlab{a}})\citenamefont{Wolf,
  McCulloch, and Schollw\"ock}}]{Wolf2014}
\bibinfo{author}{\bibfnamefont{F.~A.} \bibnamefont{Wolf}},
  \bibinfo{author}{\bibfnamefont{I.~P.} \bibnamefont{McCulloch}},
  \bibnamefont{and}
  \bibinfo{author}{\bibfnamefont{U.}~\bibnamefont{Schollw\"ock}},
  \bibinfo{journal}{Phys. Rev. B} \textbf{\bibinfo{volume}{90}},
  \bibinfo{pages}{235131} (\bibinfo{year}{2014}{\natexlab{a}}).

\bibitem[{\citenamefont{Georges et~al.}(1996)\citenamefont{Georges, Kotliar,
  Krauth, and Rozenberg}}]{Georges1996}
\bibinfo{author}{\bibfnamefont{A.}~\bibnamefont{Georges}},
  \bibinfo{author}{\bibfnamefont{G.}~\bibnamefont{Kotliar}},
  \bibinfo{author}{\bibfnamefont{W.}~\bibnamefont{Krauth}}, \bibnamefont{and}
  \bibinfo{author}{\bibfnamefont{M.~J.} \bibnamefont{Rozenberg}},
  \bibinfo{journal}{Rev. Mod. Phys.} \textbf{\bibinfo{volume}{68}},
  \bibinfo{pages}{13} (\bibinfo{year}{1996}).

\bibitem[{\citenamefont{Garc\'{\i}a et~al.}(2004)\citenamefont{Garc\'{\i}a,
  Hallberg, and Rozenberg}}]{Garcia2004}
\bibinfo{author}{\bibfnamefont{D.~J.} \bibnamefont{Garc\'{\i}a}},
  \bibinfo{author}{\bibfnamefont{K.}~\bibnamefont{Hallberg}}, \bibnamefont{and}
  \bibinfo{author}{\bibfnamefont{M.~J.} \bibnamefont{Rozenberg}},
  \bibinfo{journal}{Phys. Rev. Lett.} \textbf{\bibinfo{volume}{93}},
  \bibinfo{pages}{246403} (\bibinfo{year}{2004}).

\bibitem[{\citenamefont{Zgid et~al.}(2012)\citenamefont{Zgid, Gull, and
  Chan}}]{Zgid2012}
\bibinfo{author}{\bibfnamefont{D.}~\bibnamefont{Zgid}},
  \bibinfo{author}{\bibfnamefont{E.}~\bibnamefont{Gull}}, \bibnamefont{and}
  \bibinfo{author}{\bibfnamefont{G.~K.~L.} \bibnamefont{Chan}},
  \bibinfo{journal}{Phys. Rev. B} \textbf{\bibinfo{volume}{86}},
  \bibinfo{pages}{165128} (\bibinfo{year}{2012}).

\bibitem[{\citenamefont{Lu et~al.}(2014)\citenamefont{Lu, H\"oppner,
  Gunnarsson, and Haverkort}}]{Lu2014}
\bibinfo{author}{\bibfnamefont{Y.}~\bibnamefont{Lu}},
  \bibinfo{author}{\bibfnamefont{M.}~\bibnamefont{H\"oppner}},
  \bibinfo{author}{\bibfnamefont{O.}~\bibnamefont{Gunnarsson}},
  \bibnamefont{and} \bibinfo{author}{\bibfnamefont{M.~W.}
  \bibnamefont{Haverkort}}, \bibinfo{journal}{Phys. Rev. B}
  \textbf{\bibinfo{volume}{90}}, \bibinfo{pages}{085102}
  (\bibinfo{year}{2014}).

\bibitem[{\citenamefont{Wolf et~al.}(2014{\natexlab{b}})\citenamefont{Wolf,
  McCulloch, Parcollet, and Schollw\"ock}}]{Wolf2014b}
\bibinfo{author}{\bibfnamefont{F.~A.} \bibnamefont{Wolf}},
  \bibinfo{author}{\bibfnamefont{I.~P.} \bibnamefont{McCulloch}},
  \bibinfo{author}{\bibfnamefont{O.}~\bibnamefont{Parcollet}},
  \bibnamefont{and}
  \bibinfo{author}{\bibfnamefont{U.}~\bibnamefont{Schollw\"ock}},
  \bibinfo{journal}{Phys. Rev. B} \textbf{\bibinfo{volume}{90}},
  \bibinfo{pages}{115124} (\bibinfo{year}{2014}{\natexlab{b}}).

\end{thebibliography}

\end{document}